\documentclass[floatfix,preprint, onecolumn, showpacs,nofootinbib]{revtex4-1}
\usepackage{natbib}
\usepackage{times}
\usepackage{amssymb,amsbsy,amsmath,amsfonts}
\usepackage{graphicx}
\usepackage{float}
\usepackage{morefloats}
\usepackage{rotating}
\usepackage{srcltx}
\begin{document}

\title{Light quark mass dependence of the $D$ and $D_s$ decay constants}

\author{ L.S. Geng,$^{1,2}$ M. Altenbuchinger$^2$  and W. Weise$^2$}
\affiliation{
$^1$School of Physics and Nuclear Energy Engineering, Beihang University,  Beijing 100191,  China\\
$^2$Physik Department, Technische Universit\" at M\"unchen, D-85747 Garching, Germany}

 \begin{abstract}
We study the light quark mass dependence of the $D$ and $D_s$ meson decay constants, $f_D$ and $f_{D_s}$, using  a covariant formulation of chiral perturbation theory ($\chi$PT)
at next-to-next-to-leading order (NNLO).
Using the HPQCD lattice results for the $D$($D_s$) decay constants  as a benchmark we show that covariant $\chi$PT can describe
the HPQCD results better than heavy meson $\chi$PT (HM$\chi$PT) at both NLO and NNLO. Within the same framework, taking into account sub-leading ($1/m_Q$, with $m_Q$ the heavy quark mass) corrections to the values of the low-energy constants and employing the lattice QCD results for $g_{BB^*\pi}$, we estimate the ratio of $f_{B_s}/f_B$ to be $1.22^{+0.05}_{-0.04}$, which agrees well with the HPQCD result $1.226(26)$.
\end{abstract}

\pacs{12.39.Fe, 13.20.Fc, 14.40.Lb, 12.38.Gc}

\date{\today}

\maketitle

\section{Introduction}

The decay constants of charged pseudoscalar mesons $\pi^\pm$, $K^\pm$, $D^\pm$, $D_s^\pm$ and $B^\pm$ play an important role in our
understanding of strong interaction physics, e.g., in measurements of the Cabibbo-Kobayashi-Maskawa (CKM) matrix elements and in the search for signals of physics 
beyond the standard model (SM). At lowest order, the decay width of a charged pseudoscalar $P^\pm$ with valence quark content $q_1\bar{q}_2$ decaying into
a charged lepton pair ($\ell^\pm\nu_\ell$) via a virtual $W^\pm$ meson is given by
\begin{equation}
  \Gamma(P^\pm\rightarrow\ell^\pm\nu_\ell)=\frac{G_F^2}{8\pi} f_P^2 m^2_\ell M_P\left(1-\frac{m_\ell^2}{M^2_P}\right)|V_{q_1 q_2}|^2,
 \end{equation}
 where $m_\ell$ is the $\ell^\pm$ mass,
 $|V_{q_1 q_2}|$ is the CKM matrix element between the constituent quarks $q_1\bar{q}_2$ in $P^\pm$, and $G_F$ is the Fermi
 constant. The parameter $f_P$ is the decay constant, related to the wave function overlap of the $q_1\bar{q}_2$ pair.
 Measurements of purely leptonic decay branching fractions and lifetimes allow an experimental determination of the product $|V_{q_1q_2}f_P|$.
 A good knowledge of the value of either $|V_{q_1 q_2}|$ or $f_P$ can then be used to determine the value of the other.

These decay constants can be accessed both experimentally and through lattice Quantum Chromodynamics (lQCD) simulations. While for $f_\pi$, $f_K$, $f_D$, experimental measurements
agree well with lattice QCD calculations, a discrepancy is seen for the value of $f_{D_s}$:
The 2008 PDG average  for $f_{D_s}$ is $273\pm10$ MeV~\cite{pdg2008}, about $3\sigma$ larger than  the most precise $N_f=2+1$ lQCD result from the HPQCD/UKQCD collaboration~\cite{Follana:2007uv}, $241\pm3$ MeV. 
On the other hand,  experiments and lQCD calculations agree very well with each other on the value of $f_D$, $f_D(\mathrm{expt})=205.8\pm8.9$ MeV and $f_D(\mathrm{lQCD})=207\pm4$ MeV.
The discrepancy concerning $f_{D_s}$ is quite puzzling because whatever systematic errors have affected the lQCD calculation of $f_D$, they should also be expected for the calculation of $f_{D_s}$. 
In this context, constraints imposed by this discrepancy on new physics were seriously discussed (see, e.g., Ref.~\cite{Dobrescu:2008er}).

However, the situation has changed recently. With the new (updated) data from CLEO~\cite{Alexander:2009ux,Onyisi:2009th,Naik:2009tk} and Babar~\cite{Lee:2010qj}, together with the Belle measurement~\cite{Widhalm:2007ws},
the latest PDG average is $f_{D_s}=257.5\pm6.1$ MeV~\cite{pdg2010}\footnote{The October 2010 average from the Heavy Flavor Averaging Group (HFAG) is similar: $f_{D_s}=257.3\pm5.3$ MeV
~\cite{hfag}.}. 
The discrepancy is reduced to $2.4\sigma$. Lately the HPQCD collaboration has also updated its study of the $D_s$ decay constant~\cite{Davies:2010ip}.
By including additional results at smaller lattice spacing along with improved determinations of the lattice spacing and improved tuning of the charm and strange quark masses,
a new value for the $D_s$ decay constant has been reported\footnote{A slightly different but less precise value of $f_{D_s}=250.2\pm3.6$ MeV was obtained
in Ref.~\cite{Na:2010uf} as a byproduct from the study of the $D\rightarrow K,\ell\nu$ semileptonic decay scalar form factor by the same collaboration.}: $f_{D_s}=248.0\pm2.5$ MeV.
With the updated results from both the experimental side and the HPQCD collaboration,  the window for possible new physics in this quantity is significantly reduced~\cite{Davies:2010ip}.

An important part of the uncertainties in heavy quark lQCD simulations comes from chiral extrapolations that are needed in order to
extrapolate lQCD simulations, performed with larger-than-physical light quark masses, down to the physical point. Recent lQCD studies of the $D$ ($D_s$) decay constants, both for $N_f=2+1$~\cite{Follana:2007uv,Bazavov:2009ii} and $N_f=2$~\cite{Blossier:2009bx}, 
have adopted the one-loop heavy-meson chiral perturbation theory (HM$\chi$PT)  (including its partially-quenched and staggered counterparts) to perform chiral extrapolations. In particular,  the HPQCD collaboration has used  the standard continuum chiral expansions through first order but augmented by second- and third-order polynomial terms in $x_q=B_0 m_q/8(\pi f_\pi)^2$ where
$B_0\equiv m_\pi^2/(m_u+m_d)$ to leading order in $\chi$PT, arguing that the polynomial terms are required by the precision of the data. It is clear that the NLO HM$\chi$PT alone fails to describe its data.

HM$\chi$PT~\cite{Wise:1992hn,Yan:1992gz,Burdman:1992gh} has been widely employed not only in extrapolating lQCD simulations but
also in phenomenology studies and has been remarkably successful over the decades   (see Ref.~\cite{Casalbuoni:1996pg} for
a partial review of early applications). In Ref.~\cite{Geng:2010vw}, we have argued that a covariant formulation of $\chi$PT may be a better choice for studying heavy-meson phenomenology and lQCD simulations. This was based on the observation that the counterpart in the SU(3) baryon sector, heavy baryon $\chi$PT, converges very slowly and often fails to describe both phenomenology and lattice data (particularly the latter), e.g., in the description of the lattice data for the masses of the lowest-lying baryons~\cite{WalkerLoud:2008bp,Ishikawa:2009vc}. On the other hand, covariant baryon $\chi$PT was shown to provide a much improved description of the same data~\cite{MartinCamalich:2010fp}. Indeed, in Ref.~\cite{Geng:2010vw} we have shown that for the 
scattering lengths of light pseudoscalar mesons interacting with $D$ mesons, recoil corrections are non-negligible.
Given the important role played by $f_D$ ($f_{D_s}$) in our understanding of strong-interaction physics and the importance of chiral extrapolations in lQCD simulations,
it is timely to examine how covariant $\chi$PT works in
conjunction with the HPQCD $f_D$ ($f_{D_s}$) data.

In this letter we study the light quark mass dependence of the HPQCD $f_D$ and $f_{D_s}$ results~\cite{Follana:2007uv}\footnote{Although the HPQCD collaboration has
updated its study of the $f_{D_s}$ decay constant, it has not done the same for the $f_D$ decay constant, and therefore its $f_{D_s}/f_D$ ratio remains the same but with a slightly larger
uncertainty. For our purposes, it is enough to study the HPQCD 2007 data~\cite{Follana:2007uv}.} using a covariant formulation of $\chi$PT.
It is not our purpose to reanalyze the raw lQCD data because the HPQCD collaboration has performed a comprehensive study. Repeating such a process using
a different formulation of $\chi$PT will not likely yield any significantly different results. Instead, we will focus on their
final results in the continuum limit as a function of $m_q/m_s$, with $m_q$ the average of up and down quark masses and $m_s$ the strange quark mass. These results can be treated as quasi-original lattice data because, for chiral extrapolations, the HPQCD collaboration has used the NLO HM$\chi$PT result plus two polynomials of higher
chiral order. Therefore, any inadequacy of the NLO HM$\chi$PT should have been remedied by fine-tuning the two polynomials. Accordingly the extrapolations should be
reliable, apart from the fact that the connection with an order-by-order $\chi$PT analysis is lost.
Our present work tries to close this gap. Using the HPQCD continuum limits as a benchmark instead of the raw data not only greatly simplifies our analysis but also highlights the most important point we wish to make, namely that the covariant formulation of $\chi$PT is more suitable for chiral extrapolations of lQCD
data than the HM$\chi$PT, at least in the present case.
 
This paper is organized as follows. In Section II, we introduce the relevant effective chiral Lagrangians and calculate the Feynman diagrams
contributing to the $D$($D_s$) decay constants up to NNLO. In Section III, we show the numerical results and compare them with those of the HM$\chi$PT. We also
estimate the ratio of $f_{B_s}/f_B$ using the values of the low-energy constants (LECs) fixed in the present study and employing the lattice results for $g_{BB^*\pi}$. A short summary
follows in Section IV.

\section{Theoretical framework}
The decay constants of heavy-light pseudoscalar and vector mesons with quark content $\bar{q}Q$, with $q$ one of the $u$, $d$, and $s$ quarks and $Q$ either the $c$ or $b$ quark,  are defined by
\begin{eqnarray}
 \langle 0|\bar{q}\gamma^\mu\gamma_5 Q(0)|P_q(p)\rangle&=&-i f_{P_q} p^\mu,\\
  \langle0|\bar{q}\gamma^\mu Q(0)|P^*_q(p,\epsilon)\rangle&= &f_{P^*_q} \epsilon^\mu,
\end{eqnarray}
where $P_q$ denotes a pseudoscalar meson and $P^*_q$ a vector meson. In this convention, $f_{P_q}$  has mass dimension one and $f_{P^*_q}$ has mass dimension two~\cite{Manohar:2000dt}.  From now on, we concentrate on the charm sector, $D$, $D_s$, $D^*$, and $D^*_s$. The formalism can easily be extended to the bottom sector.

The coupling of the $D$ ($D_s$) mesons to the vacuum or to Nambu-Goldstone bosons through the left-handed current is described by the following leading chiral order Lagrangian:
\begin{equation}\label{eq:lhc}
 \mathcal{L}^{(1)}_\mathrm{source}= a\langle (c' P^*_\mu-\frac{\partial_\mu P}{m_P}) u^\dagger  \rangle,
\end{equation}
where $a$ is a normalization constant with mass dimension two, $P=(D^0,D^+,D^+_s)$, $P^*_\mu=(D^{*0},D^{*+},D^*_s)$,  $m_P$ is the characteristic mass of the $P$ triplet introduced to conserve heavy quark spin symmetry in the $m_Q\rightarrow\infty$ limit, i.e., $\mathring{m}_D$ at NLO and $m_D$ at NNLO (see Table 1), and
$u^2=U=\exp[\frac{i\Phi}{F_0}]$ with $\Phi$ the pseudoscalar octet matrix and $F_0$ their decay constant in the chiral limit.
We have introduced a dimensionless coefficient  $c'$ to distinguish the vector and pseudoscalar fields, which is 1 if heavy quark spin symmetry is exact. We need to stress that in our covariant formulation of
$\chi$PT we do not keep track of explicit $1/m_Q$ corrections that break heavy quark spin and flavor symmetry, instead
we focus on SU(3) breaking. This implies that different couplings have to be used for $D$($D^*$) and $B$($B^*$) mesons.
In the present work we only need to make such a differentiation in calculating diagram Fig.~(1d).  In Eq.~(\ref{eq:lhc}) we have therefore explicitly pointed out that $c'$ may be different from 1. In all the other places, we will simply
set $c'$ equal to 1.

The leading-order (LO) SU(3) breaking of the $D$ meson decay constants is described by 
the following next-to-leading order (NLO) chiral Lagrangians
\begin{equation}
 \mathcal{L}^{(3)}=-\frac{a}{16\pi^2 F_0^2}\left[b_D\langle (P^*_\mu-\frac{\partial_\mu P}{m_P}) \mathcal{M}\rangle+b_A
\langle P^*_\mu-\frac{\partial_\mu P}{m_P}\rangle\langle \mathcal{M}\rangle\right],
\end{equation}
where $b_D$ and $b_A$ are two LECs and $\mathcal{M}=\mathrm{diag}(m_\pi^2,m_\pi^2,2 m_K^2-m_\pi^2)$. Here
and in the following $\langle\cdots\rangle$ always denotes the trace in the corresponding flavor space. 
 
To study the NLO SU(3) breaking, one has to take into account the $DD^*$ ($D_s D^*_s$) and $DD_s$ ($D^*D^*_s$) mass splittings. 
Experimentally the $DD^*$ and $D_sD^*_s$ splittings are similar:
\begin{equation}
 \Delta_{DD^*}=141.4\,\mbox{MeV}\quad\mbox{and}\quad
  \Delta_{D_sD^*_s}=143.8\,\mbox{MeV}.
\end{equation}
Therefore in our calculation we will take an average of these two splittings, i.e., $
\Delta=(\Delta_{D D^*}+\Delta_{D_s D_s^*})/2=142.6\,\mbox{MeV}$. It should be noted that
the $DD^*$ mass splitting is of sub-leading order in the $1/m_Q$ expansion of
heavy quark effective theory. The numbers above show that SU(3) breaking of this quantity is less than $2\%$.
The mass splitting in principle can also depend on the light quark masses but we expect that the dependence
of this ``hyperfine'' splitting should be much weaker than that of  the $D$ mass\footnote{This seems to be supported by quenched lQCD calculations, see, e.g., Refs.~\cite{Guo:2001ph,Hein:2000qu}.}, $m_D$, which we discuss below.

At NLO, the following Lagrangian is responsible for generating SU(3) breaking between
the $D$ and $D_s$ masses~\cite{Geng:2010vw}:
\begin{equation}
 \mathcal{L}^{(2)}=-2c_0\langle PP^\dagger\rangle\langle\chi_+\rangle+2 c_1\langle P\chi_+ P^\dagger\rangle,
\end{equation}
which yields 
\begin{equation}\label{eq:md}
 m_D^2=m_0^2+4c_0(m_\pi^2+2m_K^2)-4c_1 m_\pi^2,
\end{equation}
\begin{equation}\label{eq:mds}
 m_{D_s}^2=m_0^2+4c_0(m_\pi^2+2m_K^2)+4c_1(m_\pi^2-2m_K^2).
\end{equation}
One may implement this mass splitting in two different ways by either using the HPQCD continuum limits on
the $D$ and $D_s$ masses~\cite{Follana:2007uv} to fix the three LECs: $m_0$, $c_0$, and $c_1$, or taking into account only
the $DD_s$ mass splitting  
\begin{equation}\label{eq:delta}
 -8 c_1 (m_K^2-m_\pi^2)=(m^2_{D_s}-m^2_D+m^2_{D_s^*}-m^2_{D^*})/2=
 \Delta_s(m_D+m_{D_s}+m_{D^*}+m_{D^*_s})/2,
\end{equation}
where we have introduced $\Delta_s\equiv m_{D_s}-m_D\approx m_{D^*_s}-m_{D^*}
\approx  (m_{D_s}-m_D + m_{D^*_s}-m_{D^*})/2$.
In the second approach, using the experimental data for $m_D$, $m_{D_s}$, $m_{D^*}$, and $m_{D^*_s}$, one obtains $c_1=-0.225$. We found that the HPQCD continuum limits on the $D$ and $D^*$ masses can be
described very well using Eqs.~(\ref{eq:md},\ref{eq:mds}). We also found that using Eqs.~(\ref{eq:md},\ref{eq:mds}) or Eq.~(\ref{eq:delta}) 
gives very similar results in our analysis of the $D$ ($D_s$) decay constants. The results shown below are obtained using Eq.~(\ref{eq:delta}) to implement
the SU(3) breaking and light quark mass evolution of the $D$ ($D_s$) masses.

In order to calculate loop diagrams contributing to the decay constants one needs to know the coupling, $g_{DD^*\phi}$,  with $\phi$ denoting
a Nambu-Goldstone boson. This 
is provided at the leading chiral order by the following Lagrangian~\cite{Geng:2010vw}:
\begin{equation}\label{eq:gddstarpi}
 \mathcal{L}^{(1)}=i g\mathring{m}_D\langle P^*_\mu u^\mu P- P u^\mu P^{*\dagger}_\mu\rangle
\end{equation}
where $u_\mu=i(u^\dagger \partial_\mu u-u\partial_\mu u^\dagger)$ and 
 $\mathring{m}_D=1972.1$ MeV, the average of $D$, $D_s$, $D^*$, and $D^*_s$ masses.
The coupling $g$ can be determined from  the $D^{*+}\rightarrow D^0\pi^+$ decay width, which
yields $g_{DD^*\pi}=0.60\pm0.07$~\cite{Geng:2010vw}. At the chiral order we are working, one can take $g_{DD^*\phi}=g_{DD^*\pi}$. If heavy quark flavor symmetry is exact, we expect $g_{BB^*\pi}=g_{DD^*\pi}$. Otherwise deviations are expected. We will come back to this later.

\begin{table*}[htpb]
      \renewcommand{\arraystretch}{1.6}
     \setlength{\tabcolsep}{0.2cm}
     \centering
     \caption{\label{table:par}Numerical values of (isospin-averaged) masses~\cite{pdg2010} and decay constants (in units of MeV) used in the present study.
 The eta meson mass is calculated using the Gell-Mann-Okubo mass relation: $m_\eta^2=(4 m_K^2-m_\pi^2)/3$.}
     \begin{tabular}{cccccccccccc}
     \hline\hline
    $\mathring{m}_{D}$     & $m_D$    &$\Delta_s$ &   $\Delta$  & $m_B$ & $\Delta_s(B)$ & $\Delta(B)$ &  $m_\pi$ & $m_K$  &    $m_\eta$   &   $f_\pi$ &  $F_0$  \\
   1972.1     & 1867.2         &  102.5   & 142.6 &
   5279.3 & 88.7 & 47.5 & 138.0  & 495.6 & 566.7 & 92.4 & $1.15 f_\pi$\\

 \hline\hline
    \end{tabular} 
\end{table*}

\begin{figure}[t]
\centerline{\includegraphics[scale=0.8]{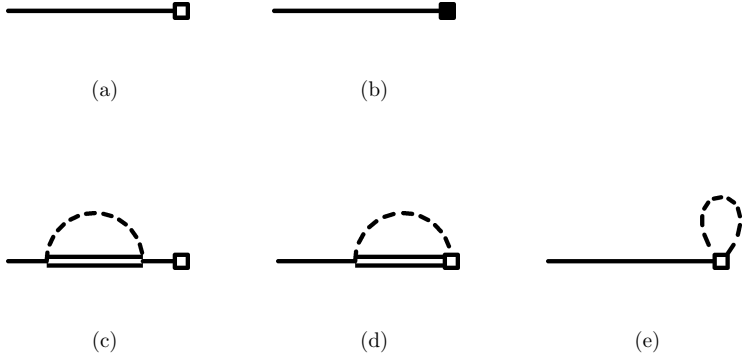}}
\caption{Feynman diagrams contributing to the calculation of $f_D$ and $f_{D_s}$
up to next-to-leading order (NLO): (a) and (b) are LO and NLO tree level
diagrams, loop diagrams (c), (d), and (e) contribute at NLO.\label{fig:diagrams}}
\end{figure}

Up to NLO, the $D$($D_s$) decay constants receive contributions from the Feynman diagrams shown in
Fig.~\ref{fig:diagrams}. Studies of these decay constants within the framework of HM$\chi$PT have a long history~\cite{Grinstein:1992qt,Goity:1992tp,Boyd:1994pa}. Here we are going to present the first covariant $\chi$PT calculation.
Insertion of the mass splittings between $D$, $D_s$, $D^*$, and $D^*_s$ in the loop diagrams shown in Fig.~\ref{fig:diagrams}
generates the NNLO contributions which are implemented in the present case by making the following replacements in the NLO results:
 \begin{equation}\label{eq:nlomass}
 m_{D_s}\rightarrow m_D+\Delta_s,\quad m_{D^*}\rightarrow m_D+\Delta,\quad\mbox{and}\quad m_{D^*_s}\rightarrow m_D+\Delta+\Delta_s,
\end{equation}
with the values of these quantities given in Table I.
It should be noted that there is no new counter-term appearing at $\mathcal{O}(p^4)$. 

Computation of the tree-level diagrams Figs.~(1a,1b) is trivial. Fig.~(1a) gives
$\hat{a}=a/m_p$ with mass dimension one for both $D$ and $D_s$ . Fig.~(1b) yields
\begin{eqnarray}
 \delta_1&=&\hat{a}\left[-\frac{1}{16\pi^2 F_0^2}\left( b_A (2 m_K^2+m_\pi^2)+b_D m_\pi^2\right)\right],\\
 \delta_2&=&\hat{a}\left[-\frac{1}{16\pi^2 F_0^2}\left(b_A (2 m_K^2+m_\pi^2)+b_D (2 m_K^2-m_\pi^2)\right)\right],
\end{eqnarray}
where $\delta_1$ is for $D$ and $\delta_2$ for $D_s$.

 Diagram Fig.~(1c) is the wave function renormalization, from which
one can calculate the wave function renormalization constants, which can be written as
\begin{equation}\label{eq:cwf}
 Z_i=\sum_{j,k} \xi_{i,j,k} \frac{d\,\phi_w(p_i^2,m_j^2,m_k^2)}{d\,p_i^2}|_{p_i^2=m_i^2},
\end{equation}
where $p_i$ denotes the four-momentum of $D$ ($D_s$), $m_i$ the mass of $D$ ($D_s$), $m_j$  the mass of $D^*$ ($D^*_s$), and $m_k$ the mass of $\pi$, $\eta$, and $K$. The coefficients $\xi_{i,j,k}$ are given in Table II. The function $\phi_w$ is defined as\footnote{To be consistent, the  product $g\mathring{m}_D$ is only appropriate for NLO. At NNLO, it has to be replaced by $g' m_D$ with $g'=g\mathring{m}_D/m_D\approx0.63$ before performing expansions in terms of $1/m_D$  either to obtain the HM$\chi$PT results or to remove the power-counting-breaking pieces. The same applies to the calculation of $C_i$ [see Eq.~(\ref{eq:ci})].}
\begin{eqnarray}
\phi_w(p_i^2,m_V^2,m_M^2)&=&\frac{(g \mathring{m}_D)^2}{4 F_0^2 m_V^2}\Big[\left(-2 m_M^2 \left(p_i^2+m_V^2\right)+\left(m_V^2-p_i^2\right){}^2+m_M^4\right)
   B_0\left(p_i^2,m_M^2,m_V^2\right)\nonumber\\
 &&+A_0\left(m_V^2\right)
   \left(-p_i^2+m_M^2-m_V^2\right)+A_0\left(m_M^2\right) \left(-p_i^2+3 m_M^2+m_V^2\right)\Big],
\end{eqnarray}
where the functions $A_0$ and $B_0$ are defined in the Appendix.

Diagram Fig.~(1d) provides current renormalization, which has
the following form
\begin{equation}\label{eq:ci}
C_i=\hat{a} c'\sum_{j,k}\xi_{i,j,k}\phi_c(m_i^2,m_j^2, m_k^2),
\end{equation}
where $\xi_{i,j,k}$ are given in Table II  with $i$ running over $D$ and $D_s$, $j$ over
$D^*$ and $D^*_s$, and $k$ over $\pi$, $\eta$, $K$. The function $\phi_c$ is defined as
\begin{eqnarray}
 \phi_c(m_i^2,m_V^2,m_M^2)&=&- \frac{(g\mathring{m}_D)m_P}{8 F_0^2 m_i^2 m_V^2}\Big[ \left(m_M^2-m_V^2\right) \left(m_i^2-m_M^2+m_V^2\right)
   B_0\left(0,m_M^2,m_V^2\right)-2
   m_i^2 A_0\left(m_M^2\right)\nonumber\\
&&+\left(-2 m_i^2
   \left(m_M^2+m_V^2\right)+m_i^4+\left(m_M^2-m_V^2\right){}^2\right) B_0\left(m_i^2,m_M^2,m_V^2\right)\Big].
\end{eqnarray}
It should be noted that $C_i$ vanishes  in NLO HM$\chi$PT but plays an important role
in covariant $\chi$PT.

Diagram Fig.~(1e) also provides current correction
\begin{equation}\label{eq:TD}
 T_i=\hat{a}\sum_{j=\pi,\eta,K} \zeta_{i,j} A_0(m_j^2)/F_0^2,
\end{equation}
with $\zeta_{i,j}$ given in Table III.

\begin{table}[htpb]
      \renewcommand{\arraystretch}{1.6}
     \setlength{\tabcolsep}{0.25cm}
     \centering
     \caption{Coefficients, $\xi_{i,j,k}$, appearing in Eqs.~(\ref{eq:cwf},\ref{eq:ci}).}
     \begin{tabular}{c|ccc|ccc}
     \hline\hline
        & \multicolumn{3}{c|}{$D^*$} &\multicolumn{3}{c}{$D^*_s$}\\\hline
        & $\pi$ & $\eta$ & $K$ & $\pi$ & $\eta$ & $K$\\\hline
  $D$   &  3 &$\frac{1}{3}$&0 & 0 &0 &2\\
  $D_s$ &  0 &0 &4 &   0 &$\frac{4}{3}$ &0\\
 \hline\hline
    \end{tabular} 
\end{table}
\begin{table*}[htpb]
      \renewcommand{\arraystretch}{1.6}
     \setlength{\tabcolsep}{0.25cm}
     \centering
     \caption{Coefficients, $\zeta_{i,j}$, appearing in Eq.~(\ref{eq:TD}).}
     \begin{tabular}{c|ccc}
     \hline\hline
        & $\pi$ & $\eta$ & $K$ \\\hline
  $D$   &  $-\frac{3}{8}$ & $-\frac{1}{24}$ & $-\frac{1}{4}$\\
  $D_s$ &  0 & $-\frac{1}{6}$ & $-\frac{1}{2}$ \\
 \hline\hline
    \end{tabular} 
\end{table*}

The total results are then
\begin{equation}
 f_i=\hat{a}(1+Z_i/2)+\delta_i+T_i+C_i.
\end{equation}

Because of the large $D$ meson masses,  $C_i$ and $Z_i$ contain so-called power-counting-breaking (PCB) terms.
As explained in detail in Ref.~\cite{Geng:2010vw} one can simply expand these functions in terms of $1/\mathring{m}_D$ at NLO
or $1/m_D$ at NNLO and then remove the PCB pieces. This procedure is in fact the same as the extended-on-mass-shell (EOMS) scheme. This scheme was first developed for baryon chiral perturbation theory~\cite{  Gegelia:1999gf,Fuchs:2003qc} and has been shown to be superior to heavy baryon $\chi$PT in a number of cases, see, e.g., Refs.~\cite{MartinCamalich:2010fp,Geng:2008mf,Geng:2009ik}. With the full results for $C_i$ and $Z_i$ given above the expansion
can easily  be performed and then one obtains $\tilde{C}_i$ and $\tilde{Z_i}$, which have a proper power-counting as prescribed in
Ref.~\cite{Geng:2010vw}. At the end one finds
\begin{equation}
 \tilde{f}_i=\hat{a}(1+\tilde{Z}_i/2)+\delta_i+T_i+\tilde{C}_i,
\end{equation}
the expression that is used in the actual calculations.
By expanding $Z_i$ and $C_i$ in terms of $1/\mathring{m}_D$ at NLO or $1/m_D$ at NNLO  and keeping the lowest order in $1/\mathring{m}_D$ ($1/m_D$) one can easily obtain the corresponding HM$\chi$PT results.

\section{Results and discussion}
Before presenting the numerical results, we should make it clear that in our present formulation of $\chi$PT we have focused on
SU(3) breaking in the context of the chiral expansions but we have not utilized explicitly heavy quark symmetry that relates the couplings of the $D$ mesons
with those of the $D^*$, $B$, and $B^*$ mesons. 

In the present case, we encounter three LECs: $a$, $b_D$, and $b_A$. At this point, light quark mass dependent lQCD results are extremely useful.
By a least-squares fit to the HPQCD results , one can fix those three LECs appearing in our calculation.  

First we treat the $D$, $D_s$, $D^*$, and $D^*_s$ mesons
as degenerate, i.e., we work up to NLO. The corresponding results are shown in Fig.~2, where the HM$\chi$PT results are obtained 
by expanding our covariant results in terms of $1/\mathring{m}_D$ and keeping only the lowest-order terms. It is clear that the covariant results (with $\chi^2$=41) 
are in much better agreement with the HPQCD continuum limits than the
HM$\chi$PT results (with $\chi^2=201$)~\footnote{It should be noted that
the absolute value of $\chi^2$ as defined here does not have a clear-cut physical meaning. It only
reflects to what extent the chiral results agree with the HPQCD extrapolations.}. This is not surprising because as we mentioned earlier the HPQCD collaboration has added second and third order polynomial terms in $x_q$ to
perform their extrapolation. Furthermore one can notice that at larger light quark masses the difference between the covariant $\chi$PT and the HM$\chi$PT results becomes larger.
This highlights the importance of using a covariant formulation of $\chi$PT in order to make chiral extrapolations if lattice simulations are
performed with relatively large light quark masses. Similar conclusions have been  reached in studying the light quark mass dependence of the lowest-lying
octet and decuplet baryon masses~\cite{MartinCamalich:2010fp}.

Taking into account the mass splittings between $D$, $D_s$, $D^*$, and $D_s^*$ as prescribed by Eq.~(\ref{eq:nlomass}) one obtains the NNLO $\chi$PT results.
Fitting them to the HPQCD extrapolations, one finds
the results shown in Fig.~3. Compared to Fig.~2, it is clear that the agreement 
between the covariant $\chi$PT results with the HPQCD extrapolations becomes even better.  Furthermore the covariant $\chi$PT results (with
$\chi^2=16$) is still visibly better than the HM$\chi$PT results (with $\chi^2=59$),
but now the difference between the covariant and the HM $\chi$PT results becomes smaller.  The three LECs in the NNLO covariant $\chi$PT have the following values:
$\hat{a}=208$ MeV, $b_D=0.318$, $b_A =0.166$.

If we had fitted the HPQCD extrapolations by neglecting the loop contributions, we would have obtained a even better
agreement ($\chi^2=9$). In Ref.~\cite{MartinCamalich:2010fp} we also found that the lattice baryon mass data could
be fitted better with the LO (linear in $m_q$) chiral extrapolation. But there we found that the NLO chiral results in fact describe the experimental data better than
the LO (linear) chiral extrapolation. This just shows that the lattice baryon mass data behave more linearly as a function of light quark masses at large light quark masses and 
chiral logarithms play a more relevant role at smaller light quark masses, as one naively expects.
 
\begin{figure}[t]
\centerline{\includegraphics[scale=1.0]{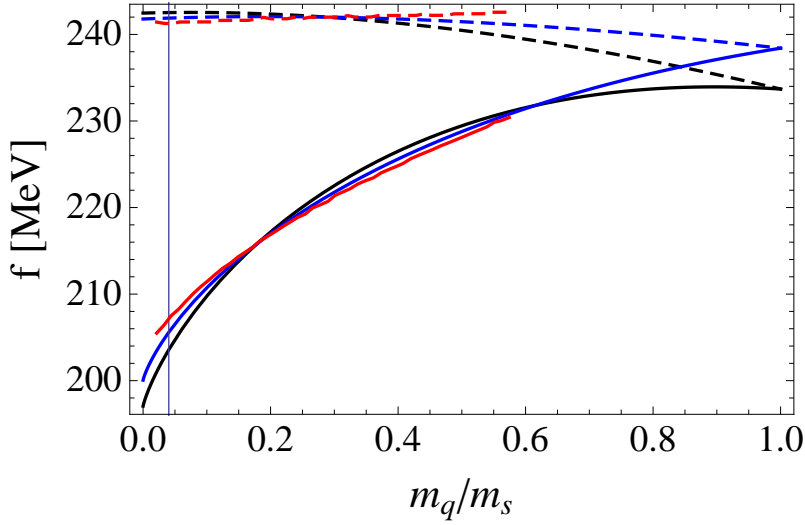}}
\caption{(Color online) Light quark mass dependence of $f_D$  (solid lines) and $f_{D_s}$ (dashed lines). 
The black lines show the results of the NLO HM$\chi$PT and the blue lines the results of the covariant NLO ChPT. 
The red lines are the continuum extrapolations of the HPQCD collaboration~\cite{Follana:2007uv}.
The ratio $r=m_q/m_s$ is related to the pseudoscalar meson masses at leading chiral order through 
$m_\pi^2=2B_0 m_s r$ and $m_K^2=B_0 m_s (r+1)$ with $B_0=m_\pi^2/(m_u+m_d)$  and
$m_q=(m_u+m_d)/2$, where $m_u$, $m_d$, and $m_s$ are the
physical up, down, and strange quark mass.
\label{fig1}}
\end{figure}

\begin{figure}[t]
\centerline{\includegraphics[scale=1.0]{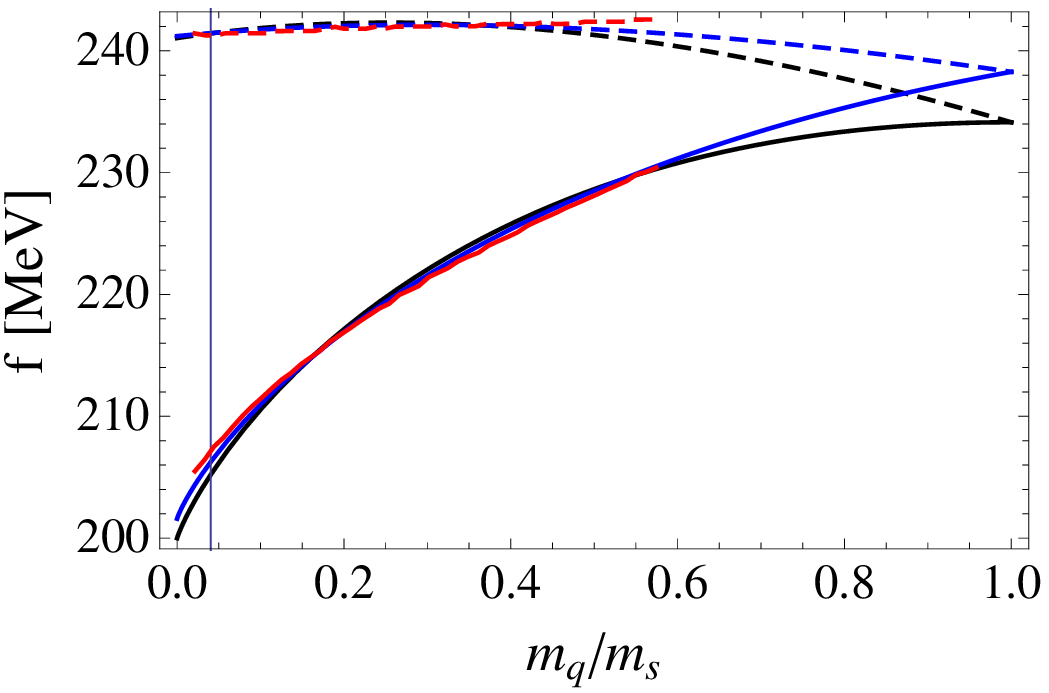}}
\caption{(Color online) Same as Fig.~\ref{fig1}, but the chiral expansions are calculated up to NNLO.\label{fig2}}
\end{figure}

\begin{figure}[t]
\includegraphics[scale=0.7]{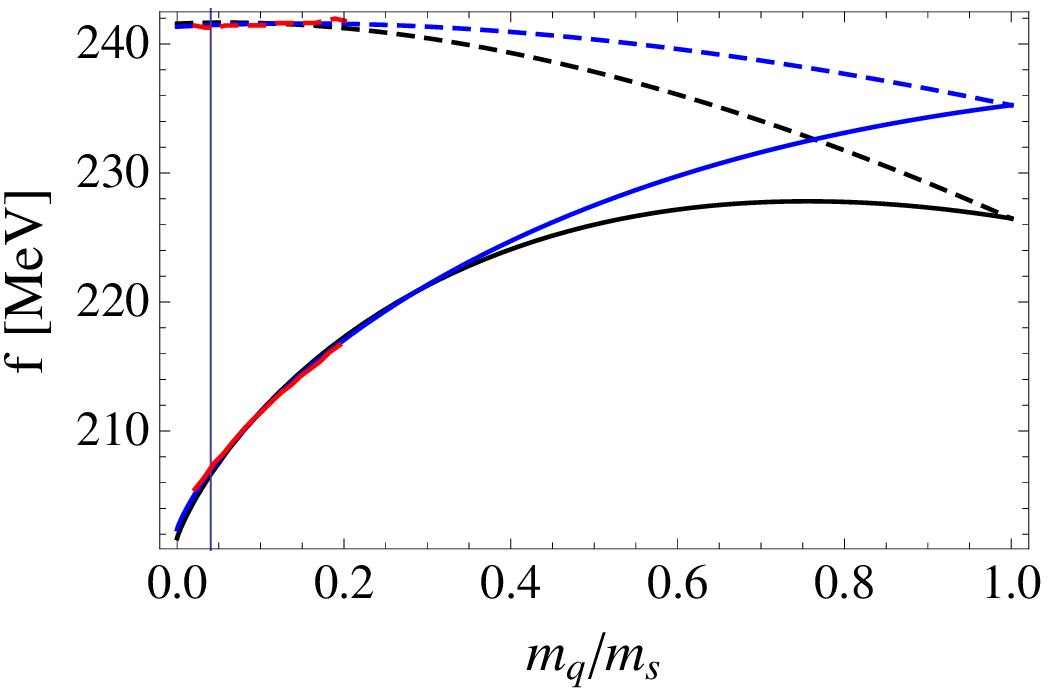}
\includegraphics[scale=0.7]{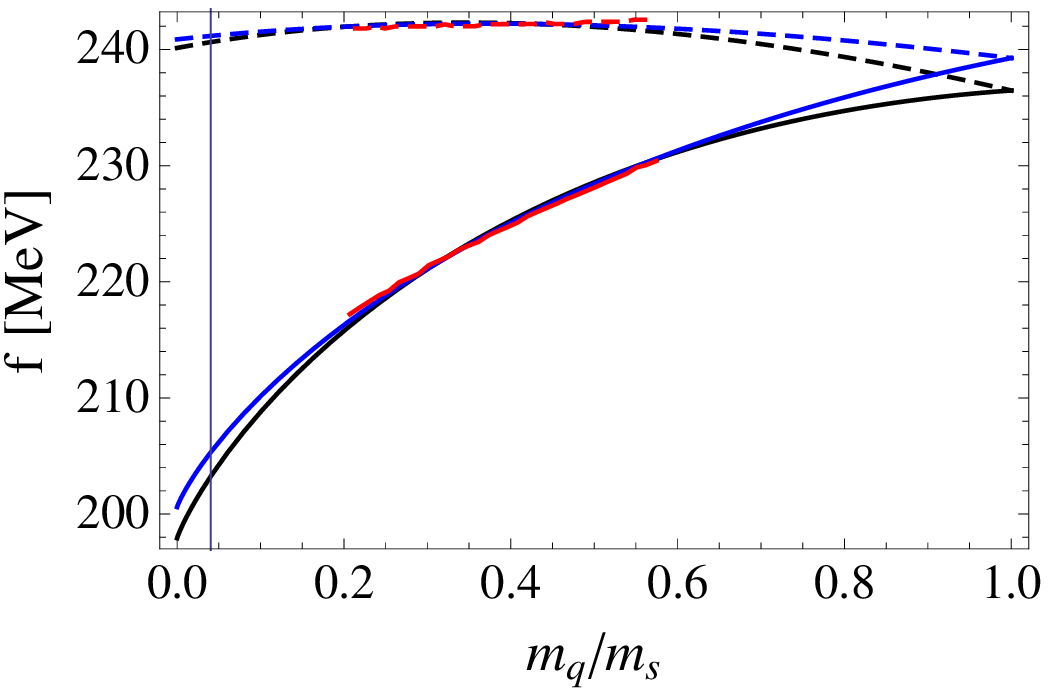}
\caption{(Color online) Same as Fig.~\ref{fig2}, but on the left panel only the lattice  extrapolations with $m_q/m_s\le0.2$ are fitted 
while on the right panel only those with $m_q/m_s>0.2$ are fitted.\label{fig2HL}}
\end{figure}

Another way of understanding the importance of chiral logarithms is to perform separate fits for lattice simulations obtained at different light quark masses. One expects that at smaller light quark masses (e.g., $m_\pi<300$ MeV) covariant $\chi$PT and HM $\chi$PT results should perform more or less similarly. On the other hand, if the light quark masses are larger, covariant $\chi$PT should be a better choice. In Fig.~4, we show the fitted results obtained from fitting the HPQCD extrapolations 
in two different regions of light quark masses, $m_q/m_s\le0.2$ (left panel) and $m_q/m_s>0.2$ (right panel). It is clearly seen that fitting lattice data with large light quark masses using the HM$\chi$PT results may give unreliable extrapolations. Here we have used the NNLO HM$\chi$PT and covariant $\chi$PT results for comparison. The difference will become even larger if the NLO $\chi$PT results are used. We should also mention that even for $m_q/m_s\le0.2$ ($m_\pi\le307$ MeV) the HPQCD extrapolations are better described by covariant $\chi$PT than by HM$\chi$PT judging from the $\chi^2$ analysis (although the difference is so small that it can hardly be appreciated by just looking at the left panel of Fig.~\ref{fig2HL}).

We have checked that our covariant results are stable with respect to
variations of certain input parameters within reasonable ranges, e.g., $m_\rho<\mu<2$ GeV and $0.53<g<0.67$, where $\mu$ is the renormalization scale and $g$ the $DD^*\pi$ coupling defined in Eq.~(\ref{eq:gddstarpi}). With our standard choice: $g=0.6$ and $\mu=1$ GeV, we have also noticed that for the NNLO covariant $\chi$PT to produce a smaller $\chi^2$ than the 
linear chiral extrapolation, $c'$ has to be larger than 1.23. If we use the quenched lQCD result, $c'=1.35\pm0.06$~\cite{Abada:1991mt}, the fit is even better\footnote{Using the results from a more recent calculation by the UKQCD collaboration~\cite{Bowler:2000xw}, one obtains $c'\approx1.18\pm0.13$, which is compatible with the result of Ref.~\cite{Abada:1991mt} but with larger uncertainties.}. 
On the other hand, our results remain qualitatively the same with either $c'=1$ or $c'=1.35$.
Therefore we have presented the results obtained with $c'=1$.

Chiral perturbation theory not only helps extrapolating lQCD simulations to the physical light quark masses. It also benefits from this process because once the values of the relevant LECs are fixed by fitting the lQCD data, $\chi$PT predicts observables involving the same set of LECs. In the present case, assuming that the $1/m_Q$ corrections to the values of the three LECs $b_D$,  $b_A$, and $g$ are small, we can calculate the ratio of $f_{B_s}/f_B$ by making the following replacements in our NNLO covariant $\chi$PT results: 
\begin{equation}
m_D\rightarrow m_B,\quad
\Delta\rightarrow \Delta (B),\quad\mbox{and}\quad
\Delta_s\rightarrow \Delta_s(B).
\end{equation}
It is found that
 deviations of $b_D$ and $b_A$ from those determined from the $D$($D^*$) mesons affect
 the $f_{B_s}/f_B$ ratio only by small amounts. Changing $b_D$ and $b_A$ by $\sim15\%$ changes $f_{B_s}/f_B$ ratio by only  about 1\%. On the other hand, the effect of $g_{BB^*\pi}$ is much larger.
 If heavy quark flavor symmetry were exact, one would have $g_{BB^*\pi}=g_{DD^*\pi}=0.6$. However, lattice QCD simulations indicate that $g_{BB^*\pi}$ is most likely smaller than $g_{DD^*\pi}$. For instance, two most recent $N_f=2$  studies give  $g_{BB^*\pi}=0.516(5)(33)(28)(28)$~\cite{Ohki:2008py} and $g_{BB^*\pi}=0.44\pm0.03^{+0.07}_{-0.00}$ ~\cite{Becirevic:2009yb}. Using 0.516 as the central value and 0.60 (0.44) as the upper(lower) bounds for
 $g_{BB^*\pi}$, we find:
\begin{equation}\label{eq:fbfbs}
f_{B_s}/f_B=1.22^{+0.05}_{-0.04},
\end{equation}
 which agrees very well with the most precise result from the HPQCD collaboration: $f_{B_s}/f_B=1.226(26)$~\cite{Gamiz:2009ku}.  The uncertainty of $\sim0.05$ does not take into account all sources of uncertainties\footnote{For instance, the small uncertainties propagated
 from the lQCD results of $f_D$ ($f_{D_s}$).}, but nevertheless it represents a reasonable
 estimate by covering a range of $g_{BB^*\pi}$ values suggested by the two recent lQCD calculations and possible $1/m_Q$ corrections to $b_D$ and $b_A$. 
 
\section{Summary and conclusions}
We have derived a covariant formulation of $\chi$PT in order to study the
$D$ and $D_s$ decay constants. To simplify the analysis, 
we have taken extrapolated lattice data (the HPQCD continuum limits) as a benchmark and focused on the light quark mass evolution of $f_D$ and $f_{D_s}$, in particular on the
SU(3) breaking pattern. We find that covariant $\chi$PT describes  the HPQCD extrapolations considerably  better than HM$\chi$PT at
a given order, although both approaches show improvement when going from NLO to NNLO. Our studies show once more that if the lattice simulations are performed with
relatively large light quark masses (e.g., $m_\pi>300$ MeV), a covariant formulation of $\chi$PT is a better choice for chiral extrapolations, particularly at low chiral orders.

Lattice QCD calculations have made remarkable progress in recent years. For ``gold plated'' physical quantities such as the decay constants studied in
this work, the overall uncertainties have been reduced to a few percent. Chiral perturbation theory plays an important role in understanding some of
the systematic errors, such as those from finite volume and extrapolations of the light quark masses to their physical values. On the other hand these precise lattice data  are also valuable
to fix the relevant LECs appearing in $\chi$PT, which can then be used to predict physical observables involving the same LECs. In the present work, we have used the HPQCD $D$ ($D_s$) data in combination with two lattice determinations of $g_{BB^*\pi}$ in order to predict the ratio $f_{B_s}/f_B=1.22^{+0.05}_{-0.04}$. This ratio turns out to be
more sensitive to the value of $g_{BB^*\pi}$ than to possible $1/m_Q$ corrections
to the two relevant LECs, $b_D$ and $b_A$.

\section{Acknowledgements}

This Work is supported in part by BMBF, the A.v. Humboldt foundation, the Fundamental Research Funds for the Central Universities (China) and by the DFG Excellence Cluster ``Origin and Structure of the Universe."
LSG thanks Prof. Christine Davies for useful and informative communications
regarding the HPQCD data.
\section{Appendix}
The functions $A_0$ and $B_0$ appearing in the calculation of the $D$ and $D_s$ meson decay constants in
the text are defined as, respectively:
\begin{equation}
 A_0(m^2)=-\frac{1}{16\pi^2}m^2\log\left(\frac{\mu^2}{m^2}\right),
\end{equation}
\begin{equation}
 B_0(p_i^2,m_1^2,m_2^2)=\left\{
\begin{array}{l}
 -\frac{1}{16\pi^2}\left[\log\left(\frac{\mu^2}{m_1^2}\right)-1\right]\quad \mbox{if}\;p_i^2=0\;\mbox{and}\;m_1=m_2\\
 -\frac{1}{16\pi^2}\left[\frac{m_1^2\log\left(\frac{\mu^2}{m_1^2}\right)-m_2^2\log\left(\frac{\mu^2}{m_2^2}\right)}{m_1^2-m_2^2}\right]\quad\mbox{if}\; p_i^2=0\;\mbox{and}\;m_1\neq m_2\\
-\frac{1}{16\pi^2}\frac{1}{{2 p_i^2}}\left[2 p_i^2 \left\{\log \left(\frac{\mu ^2}{m_2^2}\right)+1\right\}+(m_2^2-m_1^2-p_i^2) \log \left(\frac{m_1^2}{m_2^2}\right)\right.\\
    \left.+ 2 \sqrt{2 m_1^2 \left(p_i^2+m_2^2\right)-\left(m_2^2-p_i^2\right){}^2-m_1^4} \right.\\
\left.\times \left\{\tan
   ^{-1}\left(\frac{m_2^2-m_1^2-p_i^2}{\sqrt{2 m_1^2
   \left(p_i^2+m_2^2\right)-\left(m_2^2-p_i^2\right){}^2-m_1^4}}\right)
-\tan
   ^{-1}\left(\frac{m_2^2-m_1^2+p_i^2}{\sqrt{2 m_1^2
   \left(p_i^2+m_2^2\right)-\left(m_2^2-p_i^2\right){}^2-m_1^4}}\right)\right\}\right]
\end{array}\right. .
\end{equation}
In the present work, the loop results are regularized using the modified minimal subtraction scheme and, unless otherwise specified, the regularization scale $\mu$ is set at 1 GeV.

\end{document}